\documentclass[10pt,twocolumn,a4paper]{article}

\usepackage[left=20mm, right=15mm, top=15mm, bottom=20mm]{geometry}

\usepackage{subfig}
\usepackage{flushend}
\usepackage{indentfirst}
\usepackage{graphics}
\usepackage{amsmath}
\usepackage{graphicx}
\usepackage{epstopdf}
\usepackage{float}

\usepackage[font=footnotesize,labelfont=bf]{caption}
\usepackage[labelsep=period]{caption}

\graphicspath{{./figure/}}

\begin{document}

\title{\huge \textbf{A Special Physical Phenomenon --- Innate Interconnection of Space-time Points}}

\date{}

\twocolumn[
\begin{@twocolumnfalse}
\maketitle

\author{\textbf{I.\,A.\,Eganova}$^{1}$ \textbf{and} \textbf{W. Kallies}$^{2,*}$\\\\
\footnotesize $^{1}$Sobolev Institute of Mathematics, SD RAS, Novosibirsk, 630090, Russia\\
\hspace{6cm}\footnotesize $^{2}$Scientific Center of Applied
Research, JINR, Dubna, 141980, Moscow Region, Russia\\
\footnotesize $^{*}$wkallies@jinr.ru}\\\\\\

\end{@twocolumnfalse}
]

\noindent \textbf{\large{Abstract:}} \hspace{2pt} In the light of A.\,A.\,Friedman's conceptual analysis of
the World of events as a mathematical model of the physical reality
in his book ``The World as Space and Time'', {\it a priory} (innate)
interconnection of events belonging to one and the same moment of
time, which can condition the space-time metric, is considered with a
summary review of its astronomical observations by N.\,A.\,Kozyrev's
method.\\

\noindent \textbf{\large{PACS:}} \hspace{2pt} 01.55.+b; 03.30.+p; 04.20.$-$q; 97.10.$-q$\\


\vspace*{-5mm}
\noindent\hrulefill

\section{\Large{Introduction}}

Ninety years ago, in 1923, fifteen years after famous lecture
``Space and Time'' by H.\,Min\-kow\-ski, where the modern
four-dimensional mathematical model of the physical reality (the
`World of events' or `space-time') was proclaimed, A.\,A.\,Friedman's book~\cite{1} came into the world. Its title was absolutely in
the spirit of Min\-kow\-ski~--- ``The World as Space and Time''. There was
a demonstrable analysis of the basic physical categories `space' and
`time', and `motion', too, that was appreciably deepening the Way
declared by Minkowski.

Really, Friedman (1888\,--\,1925) was the only one among the contemporaries of the
special relativity creators, who penetrated into analysis of the
World of events as a mathematical model of the physical reality.
Only Friedman was then going on the cause of Minkowski who had
gone so early after his famous lecture\,---\,three and a half months
later. Probably that is why physicists, with not many
exceptions, perceived and assimilated the space-time just nominally,
without recognition of its essence, only as certain mathematical
formalism. Ipso facto A.\,Einstein's position went unheeded: it is
known that in his lectures on the essence of the theory of relativity in
Princeton in 1921, Einstein has not only high estimated Minkowski's
contribution and its function in development of the relativity,  but
also clearly stated his opinion about the World of events as the
\emph{\textbf{physical reality}}. Unfortunately, up to date there is
a diffused opinion about space-time as a convenient, but
\emph{\textbf{imaginary}} mathematical formalism.

That is why A.\,A.\,Logunov's modern analysis of the special relativity, see~\cite{Logunov}, and A.\,A.\,Sazanov's detailed examination of
Minkowski's four-dimensional World, see~\cite{Sazanov1,Sazanov2}, are very important.
That is why this Friedman's work is ever more actual. (The edition next to the
third one~\cite{2} was announced by the publishing house Harri Deutsch
Verlag (in German) in 2013.)
This book distinguishes itself by dealing with the temporal
aspect of the physical reality. In it the main
questions relating to the time\,---\,\emph{the time as a core of the
World of events, as an aspect of the physical reality}\,---\,were considered. Firstly,
Friedman showed that the core of the physical essence of the Lorentz
transform just is in idea about time as an equal in rights coordinate
of the frame of reference. In this status it does not differ from
spatial coordinates. It means that \emph{the temporal coordinate
is possessed of the physical reality  as the spatial ones and the
World of events is a mathematical model of the physical reality}.
Secondly, that is extra important, Friedman does not identify the
temporal coordinate with the spatial ones, bearing in mind their
\emph{functional} difference in universe. In his analysis the time
does not dissolve into a space many-dimensionality. That is why he
is discussing special features of the temporal coordinate and
consequently advances a new physical problem\,---\,the problem of
reinstating time as exceptional physical entity associated with
causality as the very important one and even immediately suggests
the following approach to this problem.

Causality principle \emph{``must impose known restrictions:}

1) \emph{on methods of arithmetization of the physical world at
which invariance postulate takes place; of course, one can perform
arithmetization ad libitum, but not for whatever arithmetization
the invariance postulate will take place;}

2) \emph{on properties of the geometrical four-dimensional spa\-ce to
whose interpretation the physical world corresponds; }

3) \emph{on choice of that out of the physical world coordinates to
which will be prescribe the time role} (\emph{connected with
causality principle})\emph{''}~\cite[see p.\,68]{1}.

Unfortunately,  this approach was not perfected\,---\,soon
afterwards Friedman died prematurely.

The main aim of this paper is to show that Friedman's approach to
the space-time as a mathematical model of the physical reality and
his analysis of the notion of time give a possibility to reveal a
new fundamental phenomenon\,--\, a priory, i.\,e., innate
interconnection in the World of events, that can condition
space-time metric. To show, that really it is rather not all the
same how we relate to the World of events essence.

Probably, it is fascinating to continue theoretically to design
the universe by means of own usual mathematical resources and
suppose that the physical reality has only one aspect\,---\,the
spatial one and that the space-time is a product of the space evolution
in time. But this  implies to stay in Newtonian notions about the
physical reality, which were adequate for his problems: they
correspond to the unstructured bodies having  the same internal
state, without evolution phenomena. Thereby in fact we confine
ourselves especially to the \emph{stationary} universe. This
circumstance attracted J.\,L.\,Synge's attention and he stated:
\emph{``Steeped as we are in Newtonian ideas, it is necessary to
emphasize, even ad nauseam, that space-time cannot, in general, be
split onto space and time in any invariant way''}~\cite[see p.\,154]{4}.
We ought to take into consideration that we have the semantic
extravagance if, on the one hand, we observe and admit the
universe evolution and, on the other hand, try to si\-mu\-la\-te the universe
by means of theoretical foundation adequate only for the stationary
universe.

Generally speaking, we can see fruit of the physical reality
temporal aspect ignoring, looking at the general picture that gave us
the centennial use of  space-time. It  was interpreted in detail in
A.\,K.\,Guts's review ``Hundred years of Minkowski's absolute World
of events''~\cite{5}. Volens nolens, this picture illuminates
essentially the mathematical idea brilliance and poverty of the
concrete physical achievements, if one is in fact founded on
Newtonian ideas related to the stationary physical reality and
blindly ignores the principles, fundamental impossibility of the
`space-time' separation into `space' and `time'.

That is why Friedman's conceptual analysis of the spatial aspect of
the physical reality, including the space arithmetization and the
space metric, is important at present, because it leads to the
following conclusion:

\emph{``Thus, we cannot perfectly generate the physical actions
necessary for the experimental identification of the physical
geometry in three-dimensional space;  for us these actions are as
impossible as for us is impossible the physical actions in
two-dimensional space, where it is impossible to locate our devices
and where we ourselves cannot locate. The cause of these
difficulties is \textbf{time}, without which  there is no space
and which is conditioning not the physical three-dimensional space,
but the physical four-dimensional space\,---\,the
World''}~\cite[see p.\,47--48]{1}.

If we comprehend that the World of events is the objective physical
reality, then immediately a new field of action is opened for our
investigations. (By the way, Minkowski said in the very beginning of
his lecture about them: about a strength of the views for space and
time in question and about their radical tendency.) Really, entire
\emph{terra incognita} is opened\,---\,the new aspect of objective
reality for physics. According to the logic of things, two
organically connected inter-promotional aspects of the physical
reality, the spatial and temporal, must play different parts in
universe. The temporal aspect does not copy the spatial one, it has
its functional purpose and consequently its specific properties. So,
it is interesting and necessary to investigate them.
Really, Friedman directed to such investigations, suggesting the
problem of reinstating time as exceptional physical entity
associated with the very fundamental, key property of
universe\,---\,with causality.

The Minkowski\,--\,Friedman cause was continued after 30 years by
N.\,A.\,Kozyrev
in his theoretical and experimental investigations of
physical properties of time as an aspect of the physical reality
(see \cite{6}). Actually, these investigations are returning to the time
in physics its exceptional position connected with causality. Note
that Kozyrev's study always begins with experience and terminates in
it. So, his attention to time was called by his analysis and
synthesis of numerous astronomical observational data in search of
ways of tackling the fundamental astronomical problem of the stellar
energy nature~\cite[see p.\,71--154]{6} (the detailed conceptual
analysis of this fundamental work and criticism of positions of its
adversaries (D.\,A.\,Frank-Kamenetsky and A.\,G.\,Masevich)
are in the book~\cite{7}; note that in 2005 the English
translation~\cite{8} was published). This theoretical investigation has led him to
a conclusion that it was necessary to consider `space' and `time' as
two organically, inherently interconnected aspects of the reality,
but which had essentially different purposes. He supposed that space
acted a passive part whereas time did an active one. Later the
capital books by J.\,L.\,Synge~\cite{4}, G.\,J.\,Whitrow~\cite{9},
I.\,Prigogine~\cite{10} were evidence of great importance of the
temporal aspect for natural sciences. Synge proposed that the word
`chronometry' ought to be used for denotement of \emph{``that part
of science which deals with the concept of time, with the same wide
scope which we have learned to give to the word
`geometry'\,''}~\cite[see p.\,410]{11}. Synge felt strongly that
\emph{``Euclid put us on the wrong track, so that we put space first
and time second\,---\,a very poor second indeed, for a child's study
of chronometry hardly goes beyond learning to read the face of the
clock''}~\cite[p.\,411]{11}.

From these positions, any theoretical constructions which disregard
time's own mission at the very beginning shall be simply inadequate in essence.
This statement relies on the experimental facts relating
to the peculiar physical properties of the temporal aspect~\cite{7}.
That is why in the two next sections we give our brief review of
such an experience knowledge: in the first the relation
representing mathematically the World of events physical reality is discussed,
this relation lets to talk about a priory (innate)
interconnection of events relating to one and the same moment of
time, which can condition space-time metric; in the second are
considered astronomical evidences of the World of events physical
reality. In the next article we are going to discuss
M.\,M.\,Lavrentiev's solar experiment~\cite{IW}, which
can be estimated as \emph{experimentum crucis} for Einstein's
statement: \emph{``It is neither the point in space, nor the instant
in time, at which something happens that has physical reality, but
only the event itself''}~\cite[see p.\,31]{12}.

\section{\Large{The innate events \\ interconnection}}

As it is well-known, for measurement of temporal interval between
moments of time (duration) we use special objects\,---\,processes
(motions) continuing in time. Friedman defines that motion as the
\emph{`base'}. Suppose, as a base process, some physical process
$\lambda$ is used in our clock engine and, as a tentative duration
measure, the change of its key property $\tau_{\lambda}$ is taken.
(Such property has to change monotonously from the past to the
future.) It means that if to the moment of an event $i$ corresponds
its value $\tau_{\lambda}(i)$ and to the moment of a subsequent
event $j$ ($i \prec j$) corresponds its value $\tau_{\lambda}(j)$,
\begin{equation*}\tau_{\lambda}(i) - \tau_{\lambda}(j)\equiv
\tau_{\lambda}(i,j)\quad \forall \,i\prec j\end{equation*} is the
measure of the temporal interval between events $i$ and $j$ (the
sign `$\prec$\,' denotes `is preceding').

J.\,L.\,Synge~\cite{4} right assumes that such duration measure
makes no important physical sense and the concept of time will
acquire greater concrete sense if we imply an existence of the
\emph{`standard clock'}, e.\,g., the atomic one. So that the concept
of duration is founded on the existence of special,
\emph{`standard'} processes, and immediately arises the question
about the objective choice of that standard clock.

The problem of the standard clock choice was resolved by
G.\,J.\,Whit\-row~\cite[see Ch.\,III, \S\,8]{9}. Analysing of the
standard clock problem, he pointed out an important moment of the
time measurement: \emph{the measure of a duration which consists of
two successive durations has to be equal to the arithmetical sum of
theirs}. That is because of the time additivity
\begin{equation}
t(i,j)+t(j,k)=t(i,k) \quad \forall \,\,i\prec j \prec k, \label{1}
\end{equation}
where $t(a,b)$ is the actual measure of the duration between an
event $b$ and an event $a$ preceding it.

G.\,J.\,Whitrow started from one apparent fact that by using some
tentative measure $\tau_{\lambda}$ we can find out that

\begin{equation}
\tau_\lambda (i,j)+\tau_\lambda (j,k)\neq\tau_\lambda (i,k)\quad
\forall \,\,i\prec j \prec k. \label{2}
\end{equation}
(For example, the process $\lambda$ is the radioactive decay and
$\tau_{\lambda}$ is the part of decayed atoms~\cite[see Ch.\,III,
\S\,8]{9}.) In order to make clear what differs the nonstandard
process from the standard one, G.\,J.\,Whitrow suggests for the tentative
measure for that occurs (\ref{2}) to consider the possibility of
introduction of the temporal sum $\oplus $ which shall satisfy the
condition (\ref{1}):
\begin{equation}
\tau_\lambda(i,j)\oplus \tau_\lambda(j,k)=\tau_\lambda (i,k) \quad
\forall \,\,i\prec j \prec k. \label{3}
\end{equation}

G.\,J.\,Whitrow has analytically obtained that the temporal sum
(\ref{3}) is determined by formula
\begin{equation}
\tau_\lambda(i,j)\oplus \tau_\lambda(j,k)
=\varphi_\lambda^{-1}\{\varphi_\lambda(\tau_\lambda(i,j))+\varphi_\lambda(\tau_\lambda(j,k))\} \label{4}
\end{equation}
$$
\forall \,\,i\prec j \prec k,
$$
where $\varphi_\lambda$ is the definite monotonous function
corresponding to the process $\lambda $, $\varphi_\lambda^{-1}$ is
the inverse one. In order to obtain the (\ref{4}), he used only the
well known properties of the temporal intervals:  their
commuta\-ti\-vi\-ty, i.\,e.,
$$
\tau_\lambda(i,j)\oplus \tau_\lambda(j,k) = \tau_\lambda(j,k) \oplus
\tau_\lambda(i,j)
$$
$$ \quad \forall \,\,i\prec j \prec k,
$$
and their associativity, i.\,e.,
$$
\tau_\lambda(i,j)\oplus (\tau_\lambda(j,l)\oplus \tau_\lambda(l,k)
) = (\tau_\lambda(i,j)\oplus \tau_\lambda(j,l) )\oplus
\tau_\lambda(l,k)
$$
$$
\\\quad \forall \,\,i\prec j \prec l \prec k.
$$

The (\ref{4}) indicates that every base process $\lambda$ may be
used in principle as a standard one with the aid of the measure
$$t(i,j)=\varphi_\lambda(\tau_\lambda(i,j))$$ that satisfies the condition
(\ref{1}), i.\,e., it is the actual measure of the duration between
an event $i$ and an event $j$ because of its uniqueness (within a
scale factor),  i.\,e., generally~\cite[Ch.\,III, \S\,8]{9}
\begin{equation}
t(i,j)=C_\lambda\cdot \varphi_\lambda(\tau_\lambda (i,j)), \label{5}
\end{equation}
where $C_\lambda$ is the scale factor for the base process $\lambda$.
Thereby  G.\,J.\,Whitrow  resolved the standard clock problem:
\emph{the standard clock is a clock with the additive scale}; it is
clear, if a tentative measure $\tau_\lambda$, corresponding to the
given process $\lambda$, does not satisfy the condition (\ref{1}),
we can always unambiguously (within a scale factor) image it by
means of the function $\varphi_\lambda $ on the additive mea\-su\-re~(\ref{5}).

Whitrow's approach to the standard clock problem gave us the
possibility to find out a priory (innate) interconnection of
events relating to one and the same moment of time~\cite{13}.
Really, consider the series of standard processes $\lambda, \mu,
\nu, \ldots$ and their actual duration measures
$$C_\lambda \varphi_\lambda (\tau_\lambda ),\quad C_\mu \varphi_\mu
(\tau_\mu ),\quad C_\nu \varphi_\nu (\tau_\nu ),\quad\ldots\quad ,$$
where $C_{\lambda}$, $C_{\mu}$, $C_{\nu}$, \ldots\, are scale
factors corresponding to the base processes  $\lambda, \mu, \nu,
\ldots$\,.

According to the physical reality of the temporal aspect, if we measure the
temporal interval between events $i$ and $j$ by means of these
standard clocks with various engines $\lambda, \mu, \nu, \ldots$ (of
course, in one and the same frame of reference), results of these
measurements have not to depend on the engines $\lambda, \mu, \nu,
\ldots$\,, and the following relation must hold true:
\begin{equation}
C_\lambda \varphi_\lambda (\tau_\lambda (i,j) )= C_\mu \varphi_\mu
(\tau_\mu (i,j))= C_\nu \varphi_\nu (\tau_\nu (i,j))=\ldots\,.
\label{6}
\end{equation}

The relation (\ref{6}) indicates that all base processes $\lambda$,
$\mu$, $\nu$, $\ldots$\,, which look like independent ones, proceed
not independently, but \emph{interactively}\,---\,there is an
\emph{innate interconnection} of their key characteristics
$\tau_\lambda, \tau_\mu, \tau_\nu, \ldots$\,, which is not connected
with phenomenon of ``propagation action'' in space, but conditional
by their \emph{common existence in time}. In other words, the
relation (\ref{6}) mathematically represents the space-time physical
reality.

An example of the similar innate interconnection one can see in the
physical interconnection which is hiding behind an action of Pauli
exclusion principle.

This interconnection is representing the common, one-piece
realization (``course'') of base processes in the temporal aspect,
the same is associated in philosophy with the concept about
one-piece worldwide process. The given interconnection relates to the
temporal aspect and covers events relating to one and the same
moment of time. In other words, it is an instantaneous action, a
distance-type action. Just such interconnection is able to create
the space-time metric. We are obtaining the physical solution of
Riemann's question about space metric as to space-time\,---\,about
the intrinsic cause of appearance of space-time metric; in his
paper, A.\,D.\,Alek\-san\-drov has given only philosophical
solution, see~\cite{Aleksandrov}.

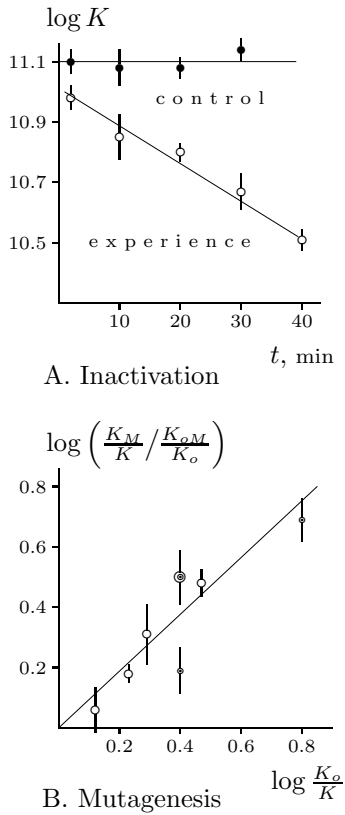
\begin{figure}
\begin{center}
\unitlength=0.8mm
\begin{picture}(70,70)
\put(15,20){ \put(0,0){\line(1,0){43}}
\multiput(10,0)(10,0){4}{\line(0,-1){1}}
\put(10,-2){\makebox(0,0)[t]{\scriptsize 10}}
\put(20,-2){\makebox(0,0)[t]{\scriptsize 20}}
\put(30,-2){\makebox(0,0)[t]{\scriptsize 30}}
\put(40,-2){\makebox(0,0)[t]{\scriptsize 40}}
\put(40,-9){\makebox(0,0){$t$, {\footnotesize min}}}
\put(0,0){\line(0,1){43}} \multiput(0,10)(0,10){4}{\line(-1,0){1}}
\put(-2,10){\makebox(0,0)[r]{\scriptsize 10.5}}
\put(-2,20){\makebox(0,0)[r]{\scriptsize 10.7}}
\put(-2,30){\makebox(0,0)[r]{\scriptsize 10.9}}
\put(-2,40){\makebox(0,0)[r]{\scriptsize 11.1}}
\put(12,-12){\makebox(0,0){A. Inactivation}}
\put(-2,47){\makebox(0,0)[l]{$\log K$}}
\put(16,34){\makebox(0,0)[l]{\scriptsize c~o~n~t~r~o~l}}
\put(5,10){\makebox(0,0)[l]{\scriptsize e~x~p~e~r~i~e~n~c~e}}
\put(2,40){\put(0,0){\circle*{1.4}}
\put(0,0.7){\line(0,1){1.3}}\put(0,-0.7){\line(0,-1){1.3}}}
\put(10,39){\put(0,0){\circle*{1.4}}
\put(0,0.7){\line(0,1){2.3}}\put(0,-0.7){\line(0,-1){2.3}}}
\put(20,39){\put(0,0){\circle*{1.4}}
\put(0,0.7){\line(0,1){1.05}}\put(0,-0.7){\line(0,-1){1.05}}}
\put(30,42){\put(0,0){\circle*{1.4}}
\put(0,0.7){\line(0,1){1.05}}\put(0,-0.7){\line(0,-1){1.05}}}
\put(2,34){\put(0,0){\circle{1.4}}
\put(0,0.7){\line(0,1){1.3}}\put(0,-0.7){\line(0,-1){1.3}}}
\put(10,27.5){\put(0,0){\circle{1.4}}
\put(0,0.7){\line(0,1){3.05}}\put(0,-0.7){\line(0,-1){3.05}}}
\put(20,25){\put(0,0){\circle{1.4}}
\put(0,0.7){\line(0,1){0.8}}\put(0,-0.7){\line(0,-1){0.8}}}
\put(30,18.5){\put(0,0){\circle{1.4}}
\put(0,0.7){\line(0,1){2.3}}\put(0,-0.7){\line(0,-1){2.3}}}
\put(40,10.5){\put(0,0){\circle{1.4}}
\put(0,0.7){\line(0,1){1.05}}\put(0,-0.7){\line(0,-1){1.05}}}
\put(0,40){\special{em:point 1}} \put(39,40){\special{em:point 2}}
\special{em:line 1,2} \put(1,35){\special{em:point 3}}
\put(39.3,11){\special{em:point 4}} \special{em:line 3,4} }
\end{picture}
\unitlength=0.8mm
\begin{picture}(70,70)

\put(15,20){ \put(0,0){\line(1,0){45}}
\multiput(10,0)(10,0){4}{\line(0,-1){1}}
\put(10,-2){\makebox(0,0)[t]{\scriptsize 0.2}}
\put(20,-2){\makebox(0,0)[t]{\scriptsize 0.4}}
\put(30,-2){\makebox(0,0)[t]{\scriptsize 0.6}}
\put(40,-2){\makebox(0,0)[t]{\scriptsize 0.8}}
\put(41,-9){\makebox(0,0){$\log\frac{K_o}{K}$}}
\put(0,0){\line(0,1){43}} \multiput(0,10)(0,10){4}{\line(-1,0){1}}
\put(-2,10){\makebox(0,0)[r]{\scriptsize 0.2}}
\put(-2,20){\makebox(0,0)[r]{\scriptsize 0.4}}
\put(-2,30){\makebox(0,0)[r]{\scriptsize 0.6}}
\put(-2,40){\makebox(0,0)[r]{\scriptsize 0.8}}
\put(12,-12){\makebox(0,0){B. Mutagenesis}}
\put(-2,47){\makebox(0,0)[l]{$\log\left(\frac{K_M}{K}\!\!\bigm/\!\!\frac{K_{oM}}{K_o}\right)$}}
\put(6,3){\put(0,0){\circle{1.4}}
\put(0,0.7){\line(0,1){3.05}}\put(0,-0.7){\line(0,-1){3.05}}}
\put(11.5,9){\put(0,0){\circle{1.4}}
\put(0,0.7){\line(0,1){0.8}}\put(0,-0.7){\line(0,-1){0.8}}}
\put(14.5,15.5){\put(0,0){\circle{1.4}}
\put(0,0.7){\line(0,1){4.3}}\put(0,-0.7){\line(0,-1){4.3}}}
\put(23.5,24){\put(0,0){\circle{1.4}}
\put(0,0.7){\line(0,1){1.55}}\put(0,-0.7){\line(0,-1){1.55}}}
\put(20,9.5){\put(0,0){\circle{1.}} \put(0,0){\circle*{0.4}}
\put(0,0.5){\line(0,1){3.25}}\put(0,-0.5){\line(0,-1){3.25}}}
\put(40,34.5){\put(0,0){\circle{1.}} \put(0,0){\circle*{0.4}}
\put(0,0.5){\line(0,1){3.}}\put(0,-0.5){\line(0,-1){3.}}}
\put(20,25){\put(0,0){\circle*{0.4}} \put(0,0){\circle{1.}}
\put(-0.1,0){\circle{1.6}}
\put(0,0.8){\line(0,1){3.7}}\put(0,-0.8){\line(0,-1){3.7}}}
\put(0,0){\special{em:point 11}} \put(42.5,40){\special{em:point
12}} \special{em:line 11,12} }
\end{picture}
\end{center}
\vspace*{-3mm}
\caption{Reaction of the cells upon the process}
\end{figure}

Naturally immediately arises a question: How can we detect such
innate interconnection in our experiment? In principle, quite
simply. We ought to pay attention to irreversible processes whereas
they relate to the base ones, to the external irreversible processes
and complex systems (i.\,e., structured systems that can be in various
internal states). In such a system there are appropriate internal
processes. That is why certain nonpower influence on the complex
system state must be observed  from an external irreversible process
(of course, under definite conditions connecting with special
properties of interconnection in question). Consider one
demonstrable example: look at the behaviour of a complex biological
system\,---\,the cells of \emph{Escherichia coly} microorganisms
under some external irreversible process, to which they are
indifferent accordingly to the conventional viewpoint. The state of the
given system is determined by means of a standard test that uses the
cells ability to form colonies on a hard agarized medium: the
viability of the cells is studied (by the number of cells capable of
reproducing), and also their state is studied by determining the
spontaneous mutation background.

As an external irreversible process to which this biological system
is indifferent accordingly to the conventional viewpoint was taken
the process associated with liquid nitrogen at room temperature.
Observations were made in special camera (it has ellipsoidal form,
the distance between focuses equals 40\,cm, the ellipsoid surface was
covered by means of aluminum foil) in order (1) to ensure a good
stability of temperature in focus where was the biological system
(in other focus was the process) and (2) to concentrate action of
the process. The open container with liquid nitrogen was located at
the bottom focus, the closed retort with the cells suspension was
located at top one. It is necessary to emphasize that temperature of
cells suspension was keeping equal
$(22,\!0\pm0,\!3)^{\circ}\mbox{C}$ during the experiment and there
is not any known action upon the cells being in anabiosis.

This experiment found out a negative reaction of the cells viability
upon the presence of liquid nitrogen at the bottom focus. Figure\,1
shows the results of this biological system testing. As we can see
in Figure\,1A, under processes in question accruing inactivation is
observed\,---\,cells are losing their viability. About the cells
viability attests the estimation their ability to reproduction: $K$
is the number of cells formed colonies in the sample,  $t$ is the
continuance of influence. In Figure\,1B we see the data showing the
efficiency of processes in question as certain mutagenic agent. The
population resistance with respect to the two antibiotics (nalidixic
acid and rifampicin) was studied: $K$ is the number of cells formed
colonies in the sample, $K_o$ is the same number in control, $K_M$
is the number of cells which are immune to the action of these
antibiotics, $K_{oM}$ is the same number in control. We see, the
number of viable cells  as compared with control decreases, but a
part of viable cells which are immune to the action of these
antibiotics as compared with control, on the contrary, increases.

In order to estimate the action in question as a mutagenic agent,
the control experiments with the known mutagens were actualized:
with the chemical and radiative one (ultraviolet). Results of this
study we are giving in Figure\,1B by means of
circles with dots inside\,---\,the action in question, as a â
mutagenic agent, does not surrender to the ultraviolet.

Thus, this experiment indicated that an external irreversi\-ble
process (to which the given biological system, accordingly to the
conventional viewpoint, is indifferent) influences in truth upon its
state. Of course, one may imagine that there is certain unknown
property of known physical interactions of which the carrier  is
liquid nitrogen and which can cause the observed cells inactivation.
That is why we ought directly to say that, when it became known that
telescope (reflector) may be used for observation of the stellar
processes influence upon the states of terrestrial complex systems,
manifold special astronomical observations were implemented~\cite{14}.
The point is that astronomical distances let to get the proof that
phenomenon in question really corresponds to the interconnection of
events relating to one and the same moment of time. Indeed, the
observations indicated that the \emph{\textbf{true}} star or stellar
system (i.\,e., location of a star or stellar system on celestial
sphere \emph{\textbf{in the observation moment}}) exerts influence
on the state of the corresponding sensor. In the time of this
observation, simultaneous visual control by sighting device (see
Figure\,4 in~\cite{14}) does not single out any celestial object
which would be projected on the sensor. Note that the refraction
phenomenon for interconnection in question is absent (it corresponds
to its belonging to the space-time temporal aspect).

\section{\Large{Astronomical evidences}}

In the astronomical observations it was unexpectedly revealed that
the used sensors react also upon the projections onto celestial
sphere of the two space-time points, $C_-$ and $C_+$ (see
Figure\,2), that relate to the observed stellar object and lie on
the corresponding light cone. Remember that we are discussing the
interconnection of events belonging to one and the same moment of
time and on the light cone lie just such. In Figure\,2 $O$ is the observer on the Earth, $C$ is the
world-line of the stellar object, $C_-$ is the event ``Apparent
stellar object'', its projection onto the celestial sphere is away
from the apparent stellar object at a distance equal to the value of
refraction for this object at the moment of observations, $C^{\ast}$
is the event ``True stellar one'', its projection onto the celestial
sphere coincides with this stellar object at the moment of
observations, $C_+$ is the event ``Symmetric stellar one'', its
projection onto the celestial sphere coincides with the position of
given object for the future, when a light signal sent from the Earth
at the moment of observations would reach it.

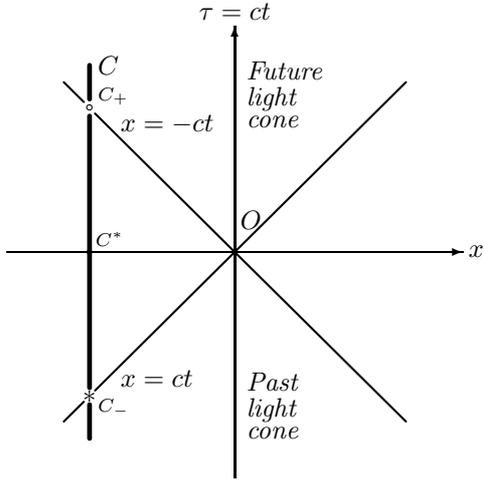
\begin{figure}
\unitlength=0.75mm
\begin{center}
\begin{picture}(90,85)(-40,-40)
\put(0,0){\vector(1,0){40}} \put(41,0){\makebox(0,0)[l]{$x$}}
\put(0,0){\line(-1,0){40}} \put(0,0){\vector(0,1){40}}
\put(0,41){\makebox(0,0)[b]{$\tau=ct$}} \put(0,0){\line(0,-1){40}}
\put(-25.5,0){\circle*{1}}
\put(-24.4,1){\makebox(0,0)[lb]{$\scriptstyle C^\ast$}}
\put(-25.5,25.5){\circle{1}}
\put(-24,26){\makebox(0,0)[lb]{$\scriptstyle C_+$}}
\put(-20,21){\makebox(0,0)[lb]{$x=-ct $}}
\put(-25.5,-25.5){\makebox(0,0){$\ast$}}
\put(-24,-26){\makebox(0,0)[lt]{$\scriptstyle C_-$}}
\put(-20,-21){\makebox(0,0)[lt]{$x=ct$}}
\put(1,4){\makebox(0,0)[lb]{$O$}} \put(2,32){\makebox(0,0)[l]{\it
Future}} \put(2,27){\makebox(0,0)[l]{\it light}}
\put(2,23){\makebox(0,0)[l]{\it cone}}
\put(2,-23){\makebox(0,0)[l]{\it Past}}
\put(2,-28){\makebox(0,0)[l]{\it light}}
\put(2,-32){\makebox(0,0)[l]{\it cone}}
\put(-24,33){\makebox(0,0)[l]{$C$}} 
\put(0,0){\special{em:point 1}} \put(30,30){\special{em:point 2}}
\put(30,-30){\special{em:point 3}}
\put(-24.5,-24.5){\special{em:point 4}}
\put(-24.5,24.5){\special{em:point 5}}
\put(-26.5,-26.5){\special{em:point 6}}
\put(-30,-30){\special{em:point 7}}
\put(-26.5,26.5){\special{em:point 8}}
\put(-30,30){\special{em:point 9}}
\put(-25.5,33){\special{em:point 10}}
\put(-25.5,27){\special{em:point 11}}
\put(-25.5,24.){\special{em:point 12}}
\put(-25.5,-24.){\special{em:point 13}}
\put(-25.5,-27){\special{em:point 14}}
\put(-25.5,-33){\special{em:point 15}} \special{em:line 1,2,0.8pt}
\special{em:line 1,3,0.8pt} \special{em:line 1,4,0.8pt}
\special{em:line 1,5,0.8pt} \special{em:line 6,7,0.8pt}
\special{em:line 8,9,0.8pt} \special{em:line 10,11,1.8pt}
\special{em:line 12,13,1.8pt} \special{em:line 14,15,1.8pt}
\end{picture}
\end{center}
\vspace{2mm} \caption {Light cone section by ($x,\tau$)-plane}
\end{figure}
Sensor's reaction on the projects onto the celestial sphere of events
$C_-$ and $C_+$ testifies to the truth of the space-time physical
reality, see above, in introduction, cited Einstein's statement
about it. In addition adequacy of Minkowski's World is confirmed.
That is why it is necessary to note the following.

1. Minkowski's World of events is in
the base for Vla\-sov\,--\,Logunov\,--\,Mest\-virishvili relativistic theory of
gra\-vi\-ta\-tion~\cite{15}.

2. Gerber's formula for Mercury's residual precession, that in due
time ensured the greatest triumph of Einstein's general relativistic
theory, can be interpreted due to the cogravitational field produced
by the apparent motion of the Sun around Mercury giving exactly the
same estimate as derived from the Schwartzschild metric in general
relativity theory~\cite{MatosTajmar}. These authors used the
generalized theory of gravitation of O.\,D.\,Jefime\-nko~\cite{16},
there Newton's theory of gravitation was developed for moving and
time-dependent gravitational systems. Generally speaking,
O.\,D.\,Jefime\-nko's theory revealed an impassable vulnerability of
Einstein's version of Gerber's formula for Mercury's residual
precession. The significant notional, semantic difficulty of the
general relativity was visualized: the mass current produced by a
moving mass distribution of density $\rho$ is $\textbf{J}=4\rho
\textbf{v}$ rather than $\textbf{J}=\rho\textbf{v}$ as would be
expected on the basis of general considerations of the mass-current
concept. Then owing to the velocity of light, as a physical
quantity, is first introduced in the general relativity when the
theory, in its limiting case, is made compatible with Poisson's
equation of the Newtonian theory of gravitation, at which time
Einstein's gravitational field equation is finally obtained, on the
one hand, and through the additional factor in formula for the
current $\textbf{J}$, on the other hand, one gets rather different,
contradictory values for the velocity of gravitation in the
different methods of linearization of Einstein's equation, see
\cite[Ch.\,20]{16}.

3. The modern experimental investigations with the aid of
coherent excitation of relativistic nuclei in a crystal (the
Japanese research teams supervised by K.\,Ko\-ma\-ki, Y.\,Yama\-zaki, and
T.\,Azuma)~\cite{17,18}, which fine-resolution confirmed the
conclusion of the special relativity concerning the relativistic
ti\-me dilation of relativistic ``clock'', actually brought in guilty
for the general relativity, as far as the corresponding energy-level
changes of relativistic nuclei undergoing tre\-mendous accelerations
inside of a crystal target were not registered in these
high-precision experiments.

The results of the Andromeda Nebula
(M\,31) observations~\cite{19} (see also~\cite[see p.\,160--166]{7} and
\cite{13}) give us a shining example:
\begin{itemize} \item{the three sensor's reactions to the
projections of the lengthy events ``Apparent M\,31'', ``True
M\,31'', and \, ``Symmetric M\,31'' (the events ``Apparent stellar
object'', ``True one'', and ``Symmetric one'' are defined above);}
\item{at the time of this observation, simultaneous
visual control by sighting device does not single out any celestial
object which would be projected on the sensor during the profile of
the projection of the events ``True M\,31'' and ``Symmetric
M\,31'';}
\item{the size of them along right ascension and along declination
corresponds to the one of the Andromeda Nebula;}
\item{the angular distances both between the projections of the
events ``Apparent M\,31'' and ``True M\,31'' and between the
projections of events ``True M\,31'' and ``Symmetric M\,31'' were
$(188\pm 2)^{''}$ along right ascension and $(34 \pm 2)^{''}$ along
declination that corresponds to the accepted data set on this
galaxy;} \item{the angular distance between the projection of the
event ``Apparent M\,31'' and the apparent M\,31 was of the order of
$23^{''}$ along declination that corresponds to the value of the
refraction of the apparent position for that celestial object at the
moment of observation\,---\,that value was equal to $23.6^{''}$;}
\item{finally, structure of all these profiles has one and the same
peculiarity\,---\,in the center of the galaxy takes place a decrease
of the sensor's reaction, it corresponds to the neutral hydrogen
distribution map in the Andromeda nebula obtained by means of the
observation data~\cite{20}, that is similar to a giant doughnut with a
hole in its center instead of the expected well known disk-shaped
distribution of stars in that galaxy.}
\end{itemize}

\section{\Large{Conclusion}}

Thus, we have the repeated astronomical observations data (of
different authors~\cite{14}) that demonstrate the physical reality
of the Minkowski's World of events\,--- we can observe the definite
points of four-di\-men\-sio\-nal World projects on celestial sphere
as images of celestial bodies. The physical reality of World of events
demands to face the facts of the absolute space-time as Minkowski at the
very beginning accentuated in his famous lecture and that
sequentially, step by step, explained and expounded Friedman in the
two first chapters of~\cite{1}, unveiling the role (function) of
time as the temporal aspect of the physical reality.

In conclusion let us put the rhetorical question: \emph{``Do we have to
know about the World of events physical reality''?} Undoubtedly, it
is necessary. Because only then may arise the understanding of that
circumstance that  it is senseless to develop ``theories of the
universe'' assuming that the physical reality has  one sole
aspect\,---\,the spatial\,---\,and ignoring the temporal and its
exceptional properties, if we wish really to know how our universe
was constructed. Remember Einstein's words from his famous
Herbert Spencer Lecture at Oxford~(1933): \emph{``Pure logical
thinking can give us no know\-led\-ge what\-so\-ever of the world of
experience; all knowledge about reality begins with experi\-ence and
terminates in it. Con\-clu\-si\-ons obtai\-ned by purely rational
processes are, so far as Reality is concerned, entirely emp\-ty...
Experience of course remains the sole cri\-te\-ri\-on of the
ser\-vi\-ce\-abi\-li\-ty of ma\-the\-ma\-ti\-cal con\-struc\-ti\-on
for phy\-sics''}, see \cite[p.\,164, 167]{Einstein}.
Friedman turned his attention to that 90 years ago,
apart from Friedman, 65 yeas ago, to such conclusion\,---\,about
\emph{urgency of temporal aspect investigations}\,---\,came Kozyrev
in consequence of his study of the stellar energy nature~\cite{8}.
Physical properties of the temporal aspect revealed by him afforded
an effective possibility to find out a priory (innate)
interconnection in the World of events considered in this paper.
Recently the properties of that phenomenon and observable effects
connected with it have been represented in detail in our
monographs~\cite{IW} and \cite{21}.


\begin{thebibliography}{99}

\bibitem{1}
Friedman A.\,A. The World as Space and Time. 2nd Edition\,--\,Nauka, Moscow, 1965; 4th Edition\,-- Publ. House ``LKI'', Moscow, 2007 (in Russian).

\bibitem{Logunov}
Logunov A.\,A. Lectures on the Theory of Relativity: Modern Analysis
of the Problem. Moscow University Press, Moscow, 1984 (in Russian).
%
\bibitem{Sazanov1}
Sazanov A.\,A. Minkowski's four-dimensional World. Nauka, Moscow,
1988 (in Russian).
%
\bibitem{Sazanov2}
Sazanov A.\,A. El universo tetradimensional de Minkowski. Editorial
Mir, Mosc\'u, 1990.
%
\bibitem{2}
Friedman A. Die Welt als Raum und Zeit.
\"Ubers., Einl. und Anm.: G. Singer,
3. bearb. Auflage. Harri Deutsch, Frankfurt am Main, 2006.
%
\bibitem{4}
Synge J.\,L. Relativity: The General Theory. North--Holland,
Amsterdam, 1960.
%
\bibitem{5}
Guts A.\,A. Hundred years of Minkowski's absolute World of events.
\emph{The search for mathematical laws of the Universe: physical
ideas, approaches, and concepts}. Academic Publ. House ``Geo'',
Novosibirsk, 2010, 13--53 (in Russian).
%
\bibitem{6}
Kozyrev N.\,A. Selected Transactions. Leningrad University Press,
Leningrad, 1991 (in Russian).
%
\bibitem{7}
Eganova I.\,A. The Nature of Space-time. Publ. House of SB RAS,
``Geo'' Branch, Novosibirsk, 2005 (in Russian).
%
\bibitem{8}
Kozyrev N. Sources of Stellar Energy and the Theory of the Internal
Constitution of Stars. \emph{Progress in Physics}, 2005, v.\,3,
61--99.
%
\bibitem{9}
Whitrow G.\,J. The Natural Philosophy of Time. Nelson and Sons,
London--Edinburgh, 1961.
%
\bibitem{10}
Prigogine I. From Being to Becoming: Time and Complexity in the
Physical Sciences.  W.\,H. Freeman and Company, San Francisco, 1980.
%
\bibitem{11}
Synge J.\,L. A plea for chronometrie. \emph{New Scientist}, 1959,
v.\,5, 410--412.
%
\bibitem{IW}
Eganova I., Kallies W. Das Sonnenexperiment von Lawrentjew. Akademiker Verlag,\hspace{0.5mm}Saarbr{\"u}cken,\hspace{0.5mm}2013.
%
\bibitem{12}
Einstein A. The Meaning of Relativity. Princeton University Press,
Princeton, N.Y., 1922.
%
\bibitem{Aleksandrov}
Aleksandrov A.\,D. About the relativity content. \emph{Einstein and
Philosophical Problems of Physics of the XX Century}. Nauka, Moscow,
1979, 117--137 (in Russian).
%
\bibitem{13}
Eganova I.\,A. The World of events reality: instantaneous action as
a con\-nec\-ti\-on of events through time. \emph{Relativity,
Gravitation, Cosmology}. Nova Science Publishers, Inc., New York,
2004, 149--162.
%
\bibitem{14}
Lavrent'ev M.\,M. and Eganova I.\,A. Kozyrev's method of
astronomical observations: information from true positions of stars,
stellar systems, and planets. \emph{Instantaneous Action at a
Distance in Modern Physics: ``Pro'' and ``Contra''}. Nova Science
Publishers, Inc., New York, 1999, 100--115.
%
\bibitem{15}
Logunov A.\,A., Mestvirishvili M.\,A. Relativistic Theory of
Gravitation. Nauka, Moscow, 1989 (in Russian).
%
\bibitem{MatosTajmar}
De Matos C.\,J., Tajmar M. Advance of Mer\-cu\-ry Pe\-ri\-he\-li\-on Ex\-plai\-ned
by Co\-gra\-vi\-ty. \emph{\,Re\-fe\-ren\-ce Fra\-mes and Gra\-vi\-to\-mag\-ne\-tism}. World Scientific, Sin\-ga\-po\-re--New Jer\-sey--Lon\-don--Hong Kong, 2001, 339--346. (http://front.math.ucdavis.edu/0304.4504)
%
\bibitem{16}
Jefimenko O.\,D. Gravitation and Cogravitation. Developing Newton's
theory of gravitation to its physical and mathematical conclusion.
Electret Scientific, West Virginia, 2006.
%
\bibitem{17}
Takabayashi Y. High precision spectroscopy of he\-li\-um-like heavy
ions with resonant coherent excitation. Doctor Thesis. Tokyo
Metropoliten Univ., Tokyo, 2001.
%
\bibitem{18}
Okorokov V.\,V. Employment of coherent excitation of relativistic
nuclei in a cristal in basic research on SRT and GRT.
\emph{Physics\,---\,Uspekhi}, 2003, v.\,46, 433--442.
%
\bibitem{19}
Kozyrev N.\,A., Nasonov V.\,V. On some properties of time detected
with the aid of astronomical observations.  \emph{Appearance of
Cosmic Factors on the Earth and Stars}. Publication of the All-Union
Astronomo-Geodesical Society of the USSR Acad. Sci., Moscow--Leningrad, 1980, 76--84 (in Russian).
%
\bibitem{20}
Roberts M.\,S. A high-resolution 21-cm hydrogen-line survey of the
Andromeda Nebula. \emph{Astrophys. J.}, 1966, v.\,144, 639--656.

\bibitem{Einstein}
Einstein A. On the Method of Theoretical Physics. \emph{Phil. Sci.},
1934, v.\,1, 63--169.
%
\bibitem{21}
Eganova I.\,A., Kallies W., Samoilow V.\,N.,  Struminsky V.\,I.
\emph{Dubna--Nauchny--Novosibirsk} Geo\-phy\-si\-cal Monitoring: The phase
trajectories of masses. Academic Publ. House ``Geo'', Novosibirsk,
2012 (in Russian).


\end{thebibliography}
\end{document}